\documentstyle[psfig]{mn}
\begin{document}
\topmargin=-0.5in
\title[Collimation of extragalactic radio jets] {Collimation of extragalactic radio jets in  compact steep spectrum and larger sources}
\author[S. Jeyakumar and D.J. Saikia]{S. Jeyakumar\thanks{E-mail: sjk@ncra.tifr.res.in} 
and D.J. Saikia\thanks{E-mail: djs@ncra.tifr.res.in} \\
National Centre for Radio Astrophysics, \\
Tata Institute of Fundamental Research, 
Ganeshkhind, Pune 411007, India \\
}
\date{}
\maketitle
\begin{abstract}
We study the collimation of radio jets in the high-luminosity Fanaroff-Riley class II
sources by examining the dependence of the sizes of hotspots and knots in the radio jets
on the overall size of
the objects for a sample of compact steep-spectrum or CSS and larger-sized
objects. The objects span a wide range in overall size from about 50 pc to nearly 1 Mpc.
The mean size of the hotspots increases with the source
size during the CSS phase, which is typically taken to be about 20 kpc, and the 
relationship flattens for the larger sources. 
The sizes of the knots in the compact as well as the larger sources
are consistent with this trend. We discuss possible implications of these trends.
We find that the hotspot closer to
the nucleus or core component tends to be more compact 
for the most asymmetric objects where the ratio of separations of the hotspots
from the nucleus, r$_d >$ 2. These highly asymmetric sources are invariably CSS objects,
and their location in the hotspot size ratio - separation ratio diagram is possibly due
to their evolution in an asymmetric environment. We also suggest that some soures,
especially of lower luminosity, exhibit an
asymmetry in the collimation of the oppositely-directed radio jets. 
\end{abstract}


\section{Introduction }
It is well established that energy is supplied more or less continuously
to the outer lobes by beams of plasma ejected from the nucleus
at relativistic speeds. The radio jets, which are
the signatures of these energy-carrying beams, 
expand rapidly with distance from the nucleus for the low-luminosity
FRI sources (Fanaroff \& Riley 1974), but are highly collimated and less
dissipative in the high-luminosity FRII objects (cf. Bridle \& Perley 1984). 
However, jets in both these classes show evidence of flaring and recollimation 
on different scales (cf. Hardee, Bridle \& Zensus 1996). The transverse widths
of the knots, $\Phi$, tend to increase with distance, $\Theta$, from the central or
nuclear component. However, the increase is not linear as would be expected in
a freely expanding jet. The spreading rate, defined to be the ratio $\Phi/\Theta$,
is usually larger close to the nucleus, which is followed by a slower spreading
rate or recollimation.
Some examples of well-studied jets in the high-luminosity FRII sources
are the ones in the 3CR quasars studied by Bridle et al. (1994, hereinafter
referred to as B94), the N 
galaxy 3C390.3 by Leahy \& Perley (1995), the radio galaxies Cygnus A (cf. Carilli 
et al. 1996) and 3C353 (Swain, Bridle \& Baum 1996). The transverse
knot widths in the inner jet in the quasar 3C351 suggest a high spreading
rate of 0.13, while the outer jet is better collimated. The jets
in the quasars 3C204, 3C263 and 3C334 exhibit faster than average spreading rate
when they are closest to the nucleus. The knots in the jets in 3C175 and 3C334 
exhibit a decrease in their transverse widths before entering the hotspots, while
the jets in the quasars 3C204 and 3C263 show evidence of flaring before entering
the hotspots (B94). 
In the N galaxy 3C390.3, the width of the jet is almost constant from 30 to 120 kpc,
it decreases till about 180 kpc and then increases again before
entering the hotspot (Leahy and Perley 1995). The width of the extended jet
in the radio galaxy 3C353 remains roughly constant beyond about 20 kpc 
(Swain et al. 1996).  Evidence of recollimation are also
seen in the small-scale nuclear jets. For example, in the 
compact steep-spectrum quasar 3C138, Fanti et al. (1989) have reported evidence of 
recollimation of the jet on scales of less than few hundred pc. The nuclear jet in 
Cygnus A has a width of about 2.2 mas over a distance ranging from
2 to 20 mas from the nucleus (Carilli et al. 1996). These observations suggest
that radio jets 
are often confined on scales ranging from the nuclear jets on the scale of parsecs
to the extended ones on scales of hundreds of kpc. The jets could be pressure
confined by the external environment or by magnetic fields around the jets
(cf. Sanders 1983; Bridle \& Perley 1984;  Appl \& Camenzind 1992, 1993a,b; 
Appl 1996; Kaiser \& Alexander 1997).

In this paper, we concentrate on the hotspots and radio knots in the jets and 
use their sizes over a large
range of source sizes to investigate the collimation of jets in the compact
steep spectrum and larger sources. The hotspots indicate the
Mach disks where the jets terminate and  subtend small angles 
in the radio cores suggesting the high degree of collimation of the jets in
these sources (Bridle \& Perley 1984; B94; Fernini, Burns \& Perley 1997, 
hereinafter referred to as F97).
Inclusion of hotspots enables us to study the collimation of jets for a large sample of sources
independent of whether a jet has been detected and its transverse width determined.
However, the jet momentum may be spread over a larger area than the cross-section of
the jet itself due to the dentist-drill effect discussed by Scheuer (1982),
where the end of the jet wanders about the leading contact surface drilling
into the external medium at slightly different places at different times. 
Recent simulations of 3D supersonic jets suggest that cocoon turbulence
drives the dentist-drill effect (Norman 1996).

Over the last few years the sizes of hotspots have been determined for 
samples of compact steep-spectrum radio sources  using largely VLBI and MERLIN 
observations, as well as for the larger objects using the VLA. Although 
there is no well-accepted definition of a hotspot (cf. Laing 1989; Perley 1989),
we are interested in features which mark the collimation of jets 
and have used the following empirical definition. The hotspots
are defined to be the brightest features in the lobes located further from
the nucleus than the end of any jet, and in the
presence of more extended diffuse emission these should be brighter by atleast a factor
of about 4 (cf. B94). In the presence of multiple hotspots, only the
primary hotspot has been considered. The jets which we consider have been well-mapped
and follow the defining criteria suggested by Bridle \& Perley (1984). 
We have used the information on the sizes of the hotspots and the widths of the knots
in the jets to study the collimation and expansion of the radio jets over a
wide range of angular and linear scales. 
The overall linear sizes of our objects are spread over about 5 orders of magnitude
ranging from about 50 pc to nearly 1 Mpc.

\begin{table*}

\caption{The sample of largely CSS and GPS sources}
\begin{tabular}{l l l l l l r r r r l r l l l }
\hline
IAU name & Alt. name & Id & Sample & z & St & LAS & Size & n$_b$ & \multicolumn{4}{c}{Hotspot sizes} & r$_d$ & Ref\\
     &   &   & &  &  & &  &      & \multicolumn{2}{c}{Hotspot 1} & \multicolumn{2}{c}{Hotspot 2} & & \\
     &   &   &  & &  & arcsec & kpc & &$mas$ & $mas$ & $mas$ & $mas$ & & \\
(1) & (2) & (3) & (4) & (5 ) & (6) & (7) & (8) & (9) & (10) & (11) & (12) & (13) & (14) & (15) \\
\hline
0108+388 &         OC+314 &  G &   O,T &0.669 &  T & 0.0058 & 0.054 & 12 &  0.47 &0.23 &  0.53 &0.39 &1.34 &1  \\
0138+138 &          3C 49 &  G &   O,S &0.621 &  T &      1 &     9 & 14 &   100 &  40 &    20 &  10 &2.42 &2  \\
0221+276 &          3C 67 &  G &   O,S &0.310 &  T &   2.47 &    15 & 41 &   160 &  60 &   115 &  98 &2.45 &2  \\
0223+341 &       4C+34.07 & Q? &   O,S &        &  D &  0.546 &        & 18 &$\star$   30 &  26 &$\star$ $<$  30 &$<$  30 &        &4  \\
0404+768 &        4C76.03 &  G & O,S,T &0.599 & T? &   0.11 &  0.97 & 11 &$\star$   10 &  10 &$\star$   10 &   8 &1.64 &5,4  \\
0428+205 &          OF247 &  G &   O,S &0.219 & T? &  0.212 &     1 & 21 &$\star$   17 &  10 &$\star$   11 &   7 &4.68 &4  \\
0518+165 &         3C 138 &  Q &   O,S &0.759 &  T &  0.615 &     6 & 12 &$\star$   71 &  37 &$\star$   64 &  25 &1.72 &18  \\
0538+498 &         3C 147 &  Q &   O,S &0.545 &  T &   0.64 &   5.4 & 32 &    &    &$\star$   70 &  49 &2.50 &17  \\
0707+689 &       4C+68.08 &  Q &     S &1.139 &  T &   1.18 &    13 & 19 &   170 & 100 &    &    &1.24 &2  \\
0710+439 &         OI+417 &  G &   O,T &0.518 &  T &  0.025 &  0.21 & 25 &  1.08 &0.63 &  0.48 &0.13 &1.12 &1  \\
0740+380 &         3C 186 &  Q &   O,S &1.063 &  T &   2.19 &    24 & 26 &$\star$  311 & 145 &$\star$   94 &  50 &1.29 &9  \\
0758+143 &         3C 190 &  Q &   O,S &1.197 &  T &    2.6 &    30 & 76 &$\star$  101 &  76 &$\star$   89 &  76 &1.61 &9  \\
0802+103 &         3C 191 &  Q &     S &1.956 &  T &    4.6 &    59 & 42 &$\star$  248 & 154 &$\star$  160 &  85 &1.75 &6,11,2  \\
0858+292 &       3C 213.1 &  G &     S &0.194 &  D &   5.95 &    26 & 20 &   300 & 200 &   200 & 200 &        &3  \\
1019+222 &         3C 241 &  G &   O,S &1.617 &  T &   0.81 &    10 & 12 &    60 &  30 &    60 &  30 &1.33 &2,8  \\
1031+567 &         OL+553 &  G &   O,T &0.459 &  D & 0.0334 &  0.26 & 24 &$\star$  1.7 & 1.5 &$\star$  1.5 & 1.1 &        &1  \\
1117+146 &       4C+14.41 &  G &   O,S &0.362 &  D &   0.08 &  0.54 & 11 &    15 &   7 &    20 &   6 &        &2  \\
1122+195 &         3C 258 &  G &     S &0.165 &  D &    0.1 &  0.38 & 14 &    15 &  10 &    15 &   8 &        &2  \\
1143+500 &         3C 266 &  G &     S &1.275 &  D &   4.16 &    49 & 14 &   500 & 100 &   700 & 300 &        &3  \\
1203+645 &       3C 268.3 &  G &   O,S &0.371 &  T &   1.36 &   9.3 & 13 &   190 & 110 &   118 &  81 &3.06 &12,2  \\
1225+368 &          ON343 &  Q &   O,S &1.974 & T? &  0.055 &  0.71 & 18 &$\star$ $<$   7 & 3.2 &$\star$ $<$   7 &   4 &1.21 &4  \\
1244+492 &       4C+49.25 &  G &     S &0.206 &  T &   2.64 &    12 & 33 &   125 &  90 &   230 & 160 &1.63 &2  \\
1250+568 &       3C 277.1 &  Q &   O,S &0.321 &  T &    1.7 &    11 & 57 &    &    &$\star$   50 &  37 &2.55 &2  \\
1323+321 &       4C+32.44 &  G &   O,S &0.370 &  D &  0.055 &  0.37 & 11 &$\star$   10 &   6 &$\star$    9 &   3 &        &4  \\
1328+307 &         3C 286 &  Q &   O,S &0.849 &  T &    3.2 &    33 & 11 &   600 & 300 &   200 & 200 &3.36 &3,2  \\
1402+660 &       4C+66.14 & EF &     S &        &  T &   0.17 &        & 24 &    &    &    26 &  12 &1.37 &2  \\
1416+067 &         3C 298 &  Q &   O,S &1.439 &  T &    1.5 &    18 & 19 &   130 & 130 &    90 &  90 &2.70 &8  \\
1419+419 &         3C 299 &  G &     S &0.367 &  T &   11.5 &    78 & 10 &$\star$  824 & 779 &$\star$  824 & 576 &3.21 &16,12  \\
1447+771 &       3C 305.1 &  G &   O,S &1.132 & T? &   2.34 &    27 & 16 &   149 & 119 &    89 &  46 &2.40 &8,2  \\
1517+204 &         3C 318 &  G &   O,S &0.717 & T? &  0.724 &     7 & 21 &$\star$   72 &  48 &    90 &  35 &1.41 &9  \\
1518+047 &       4C+04.51 &  G &     O &1.296 &  D &  0.135 &   1.6 & 23 &  $<$ 3.6 &$<$ 3.6 &   1.7 & 1.7 &        &13,14  \\
1607+268 &          CTD93 &  G &   O,S &0.473 &  D &   0.05 &  0.39 & 30 &  0.51 &0.51 &  4.95 &2.23 &        &10  \\
1946+708 &            TXS &  G &     T &0.101 &  T &  0.031 & 0.079 & 24 &  5.04 &0.76 &  0.93 &0.69 &1.00 &15  \\
2128+048 &            PKS &  G &     O &0.990 &  T &  0.035 &  0.38 & 12 &   3.2 & 2.9 &   3.2 & 1.6 &1.44 &7  \\
2210+016 &       4C+01.69 & G? &     O &        & T? &  0.075 &        & 17 &   5.1 & 2.9 &    &    &1.85 &7  \\
2252+129 &         3C 455 &  Q &   O,S &0.543 &  D &   3.12 &    26 & 10 &$\star$  871 & 673 &   800 & 400 &        &3  \\
2323+435 &          OZ438 &  G &     S &0.145 & T? &   1.61 &   5.6 & 20 &   140 &  90 &   110 &  90 &2.04 &2  \\
2342+821 &             S5 &  Q &   O,S &0.735 & T? &   0.16 &   1.6 & 13 &$\star$   17 &  15 &$\star$   17 &  12 &1.39 &4  \\
2352+495 &         OZ+488 &  G &   O,T &0.237 &  T &  0.049 &  0.25 & 55 &  2.06 &1.26 &   1.2 &0.83 &1.05 &1  \\
\hline
\end{tabular}
\\
References: 
1.  Taylor et al. 1996; 
2.  Sanghera et al. 1995; 
3.  Spencer et al. 1989; 
4.  Dallacassa et al.  1995; 
5.  Polatidis et al. 1995; 
6.  Cawthorne et al. 1986; 
7. Stanghellini et. al. 1997; 
8. van Breugel et al. 1992; 
9. Spencer et al 1991; 
10. Fey \& Charlot 1997; 
11. Pearson, Perley \& Readhead 1985; 
12. Rendong et al. 1991b; 
13. Mutel, Hodges  \& Phillips 1985; 
14. Phillips \& Mutel 1981; 
15. Taylor, Vermeulen \& Pearson 1995;  
16. Liu, Pooley \& Riley 1992; 
17.  Simon et al. 1990;  
18.  Akujor et al. 1993. 
\end{table*}

\begin{table*}

\caption{The sample of large 3CR sources}
\begin{tabular}{l l l l r r r r l r l l l }
\hline
IAU name & Alt. name & Id &  z & LAS & Size & n$_b$ & \multicolumn{4}{c}{Hotspot sizes} & r$_d$ & Ref\\
     &   &  &  &  &  &      & \multicolumn{2}{c}{Hotspot 1} & \multicolumn{2}{c}{Hotspot 2} & & \\
     &   &    &  & arcsec & kpc & &$mas$ & $mas$ & $mas$ & $mas$ & & \\
(1) & (2) & (3) & (4) & (5 ) & (6) & (7) & (8) & (9) & (10) & (11) & (12) & (13) \\
\hline
0013+790 &         3C 6.1 &  G & 0.840 &    26 & 266.3 & 13 &  1300 &$<$ 900 &  $<$ 900 &$<$ 900 &        &2  \\
0017+154 &           3C 9 &  Q & 2.012 &    14 & 181.1 & 39 &   900 & 280 &   620 & 360 &1.52 &9  \\
0040+517 &          3C 20 &  G & 0.174 &    53 & 211.4 & 26 &  1700 &2000 &  $<$1000 &1700 &        &5,1  \\
0048+509 &          3C 22 &  G & 0.936 &    24 & 255.8 & 12 &  $<$1000 &1300 &  $<$1000 &1300 &        &1  \\
0107+315 &          3C 34 &  G & 0.690 &    48 & 453.4 & 24 &  1700 &2300 &    &    &        &1  \\
0229+341 &        3C 68.1 &  Q & 1.238 &    53 & 616.5 & 15 &  1500 &1400 &    &    &        &5,9  \\
0605+480 &         3C 153 &  G & 0.277 &   7.1 &  39.9 & 27 &   110 &  70 &   190 &  80 &2.00 &8  \\
0702+749 &       3C 173.1 &  G & 0.292 &    60 & 349.6 & 29 &  2500 &1800 &  3600 &2400 &1.18 &1  \\
0809+483 &         3C 196 &  Q & 0.871 &   5.5 &  57.1 & 22 &   700 & 500 &   500 & 150 &1.32 &7  \\
0833+654 &         3C 204 &  Q & 1.112 &  31.1 & 350.7 & 16 &  1500 &$<$1000 &  2000 &$<$1000 &        &2,5  \\
0835+580 &         3C 205 &  Q & 1.534 &    16 & 196.3 & 20 &  $<$ 300 &$<$ 400 &  $<$ 400 &$<$ 500 &        &5,2  \\
0850+140 &         3C 208 &  Q & 1.110 &    14 & 157.8 & 38 &   260 & 210 &   240 & 210 &1.30 &9,5  \\
0926+793 &       3C 220.1 &  G & 0.610 &    30 & 267.8 & 15 &  $<$1000 &2000 &    &    &        &1  \\
1009+748 &       4C+74.16 &  G & 0.200 &    40 & 177.6 & 19 &   500 & 700 &  2500 &2000 &1.41 &6  \\
1030+585 &       3C 244.1 &  G & 0.428 &    53 & 392.6 & 23 &  1000 &1200 &  1000 &2000 &1.15 &1,5  \\
1100+772 &       3C 249.1 &  Q & 0.311 &    23 & 139.8 & 11 &  1300 &1000 &  1300 &$<$1000 &        &2  \\
1108+359 &         3C 252 &  G & 1.105 &    60 & 675.3 & 28 &  $<$1000 &1700 &    &    &        &1  \\
1111+408 &         3C 254 &  Q & 0.734 &  13.2 & 128.0 & 19 &  $<$ 200 &$<$ 300 &  $<$ 200 &$<$ 300 &        &5,2  \\
1137+660 &         3C 263 &  Q & 0.656 &  44.2 & 408.2 & 22 &  $<$1000 &$<$1000 &    &    &        &2  \\
1142+318 &         3C 265 &  G & 0.811 &    78 & 788.0 & 35 &  2200 &1500 &  1100 &2100 &1.50 &1,5  \\
1157+732 &       3C 268.1 &  G & 0.970 &    46 & 496.3 & 22 &  $<$1000 &$<$1000 &  $<$1000 &$<$1000 &        &1,5  \\
1254+476 &         3C 280 &  G & 0.996 &  12.9 & 140.4 & 19 &  $<$ 200 &$<$ 300 &  $<$ 200 &$<$ 300 &        &5,2  \\
1533+557 &         3C 322 &  G & 1.681 &    33 & 412.9 & 14 &  1100 &1600 &  1800 &$<$1000 &        &1  \\
1609+660 &         3C 330 &  G & 0.550 &    62 & 526.0 & 30 &  1100 &$<$1000 &  3300 &$<$1000 &        &1,5  \\
1627+444 &         3C 337 &  G & 0.630 &    43 & 389.7 & 21 &  $<$1000 &$<$1500 &  1300 &$<$1500 &        &1  \\
1658+471 &         3C 349 &  G & 0.205 &    82 & 371.0 & 33 &  1600 &1700 &    &    &        &1,5  \\
1704+608 &         3C 351 &  Q & 0.371 &    64 & 435.3 & 29 &  1500 &1000 &    &    &        &3,9  \\
1723+510 &         3C 356 &  G & 1.079 &    75 & 837.9 & 29 &  $<$1000 &1900 &  1600 &$<$1500 &        &1  \\
1832+474 &         3C 381 &  G & 0.161 &    69 & 258.9 & 26 &  $<$1500 &$<$1000 &  6000 &3000 &        &3,5  \\
2104+763 &       3C 427.1 &  G & 0.572 &  23.1 & 199.9 & 11 &  5700 &2300 &  2900 &1500 &1.01 &2  \\
2120+168 &         3C 432 &  Q & 1.805 &  13.4 & 170.0 & 36 &   310 & 160 &   260 & 230 &1.46 &9  \\
2352+796 &       3C 469.1 &  G & 1.336 &    74 & 878.4 & 37 &  $<$ 900 &1300 &  $<$ 900 &2200 &        &4  \\
\hline
\end{tabular}
\\
References: 1.  Jenkins, Pooley \& Riley 1977; 
2.  Pooley \& Henbest 1974; 
3.  Riley \& Pooley   1975; 
4.  Longair  1975; 
5.  Laing  1981; 
6.  Laing, Riley \& Longair  1983; 
7.  Lonsdale \& Morison  1983; 
8.  Hardcastle et al. 1998; 
9.  Bridle et al. 1994.             
\end{table*}

\section{Sample of sources}
Our sample has been chosen from well-defined samples of compact and
larger sources which have been observed with high angular resolution.
We have compiled our sample of compact sources from 
the following samples of CSS and Gigahertz Peaked Spectrum or 
GPS sources: (i) the sample of 67 sources listed by O'Dea (1998) and O'Dea \& Baum (1997) 
which is based on the 
Fanti et al. (1990) CSSs and Stanghellini et al. (1990, 1996) GPS objects; 
(ii) the Sanghera et al. (1995) compilation of 62 objects, which consists of all sources 
from the complete samples of 3C (Fanti et al. 1990), PW (Peacock \& Wall 1982) 
and a Jodrell-Bank sample; and (iii) 
the 7 confirmed compact symmetric objects or CSOs listed by Taylor,
Vermeulen \& Pearson (1995) and Taylor, Readhead \& Pearson (1996).
Our resulting combined sample consists of 86
objects, because there are many sources which are common to the
above lists, and have been counted only once in our combined list. 
Of these 86, we have considered those which have a well-defined double-lobed
structure and where the hotspot sizes are available or could be determined. Sources with
a complex and highly distorted structure have been excluded. Typical examples of such sources are 
3C48 (Wilkinson et al. 1991) and 3C119 (Ren-dong et al. 1991a). In addition, we have
excluded the source 0319+121 which has been listed as a GPS object by Stanghellini
et al. (1990), but single-epoch observations of the nucleus from 1.4 to 15 GHz show
that it has a flat radio spectrum (Saikia et al. 1998). 
A few of our sources, such as 0108+388 (Baum et al. 1990), 
have extended emission in addition to the compact double or triple structure 
closer to the nucleus. The extended emission must be due to 
earlier periods of activity, with the luminosity and size of the component being
governed by ageing and expansion of the component. On the other hand, 
the compact structure closer to the nucleus represent more recent activity with
the size of the hotspots indicating the degree of collimation of the jets. Hence,
in such cases we have considered only the compact double or triple structure.

The sample which consists of 58 objects at this stage 
has been further restricted to those which have been observed with at least 10
resolution elements, n$_b$, along the main axis. We also focus on the subsample which
has been observed with n$_b \geq$ 20, for a more reliable estimation of the 
hotspot parameters. In our entire sample only 6 lobes have no hotspots which meet
our defining criteria. Excluding these lobes does not affect any of the conclusions
presented in this paper.
Sources which were earlier classified as CSS objects but were later found to be
of larger dimensions have not been excluded from the sample.
Our final sample of largely CSS and GPS objects consists of 39 sources, 9 of which
are $>$20 kpc. This sample is listed in
Table 1, which is arranged as follows: columns 1 and 2: source name and an 
alternative name; column 3: optical identification where G denotes a galaxy, 
Q a quasar and EF an empty field; 
column 4: the sample where O, S and T  denote O'Dea \& Baum (1997) and O'Dea (1998), 
Sanghera et al. (1995) and Taylor et al. (1995, 1996) respectively;
column 5: redshift; column 6: the structural classification of the source where D
denotes a double, T a triple with a radio core, and T? a triple with a possible
core component; column 7: the angular separation of the hotspots on
opposite sides of the nucleus, expressed in arcsec; column 8: 
the corresponding linear size in kpc
in an Einstein-de Sitter Universe with H$_o$=50 km s$^{-1}$ Mpc$^{-1}$;
column 9: the number of resolution elements, n$_b$, along the longest axis of the 
source, which is the largest angular size of the radio source divided by the size of
the restoring beam along the axis of the source; 
columns 10 and 11: the major and minor axes of one hotspot in mas;
columns 12 and 13: the
major and minor axes for the other hotspot in mas; column 14 : ratio of the separation
of the farther component from the radio core to the nearer one for those classified as
T or T?; column 15: references for radio structure. For
sources with a radio core, the size of the hotspot farther from the nucleus
is listed in columns 10 and 11. The hotspot sizes refer to the full width at half
maximum. For some of the images where the authors quote the size of the 
entire component from the lowest reliable contour, we have estimated the sizes of
the hotspots from the images. These have been marked with an asterisk in the Table.

For comparison with larger sources we 
consider those which have been observed with a similar number of resolution
elements, n$_b$, along the main axis of the source.
The comparison sample has
been compiled from the complete sample of 3CR sources (Laing, Riley \&
Longair 1983) and consists of the FRII sources which have been observed with 
n$_b$ between about 10 and 40. These limits were chosen
so that the distribution of n$_b$ is similar to the CSS and GPS objects.
For the 3CR sources observed with the Cambridge 5-km telescope
we have confined ourselves to sources above a declination of 30$^\circ$ so
that the beams are not very elliptical. This sample is listed in Table 2
which is arranged similarly to Table 1 except for the following differences.
These are all triples from the 3CR sample, and hence the sample column and 
structual information have been omitted. 

In addition, we discuss briefly the collimation of jets using sources 
which have been observed with a larger number of resolution elements along their axes.
These have been been observed with high resolution
and sensitivity with the Very Large Array (VLA) by B94,
Fernini et al. (1993), F97, Black et al. (1992), Leahy et al. (1997), 
and Hardcastle et al. (1997, 1998). These observations give us estimates
of the sizes of hotspots in larger sources. B94 have reported observations of 13
quasars while F97 have listed hotspot sizes for 9 galaxies. Hardcastle et al.
(1998, hereinafter referred to as H98) have summarized the properties of jets, 
cores and hotspots in the sample
of FRII galaxies with redshift $<$0.3 which have been observed by the Cambridge group.
In addition to the hotspots we have also considered the sizes of the knots in the jets listed in 
Table 9 of B94 and of the knots in the CSOs observed by Taylor et al.  (1995, 1996). 

\begin{figure*}
\vspace{-5.8in}
\vbox{
\hbox{
\hspace{-0.5in}
\psfig{figure=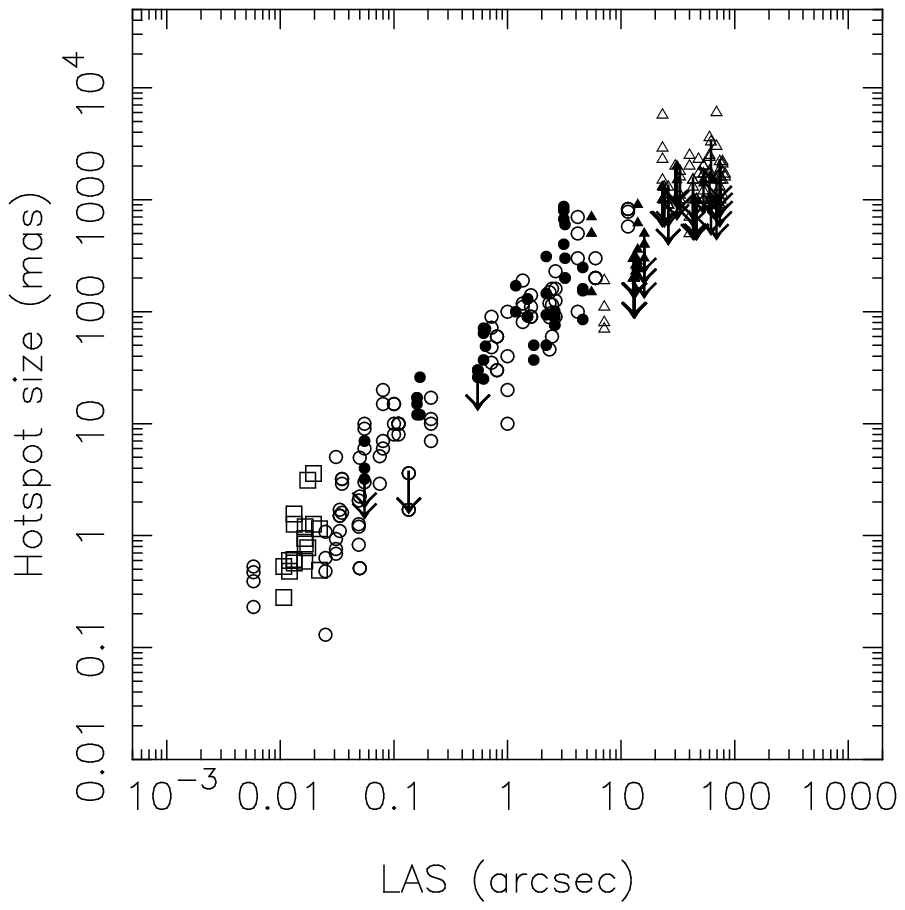,width=7.1in}
\hspace{-3.6in}
\psfig{figure=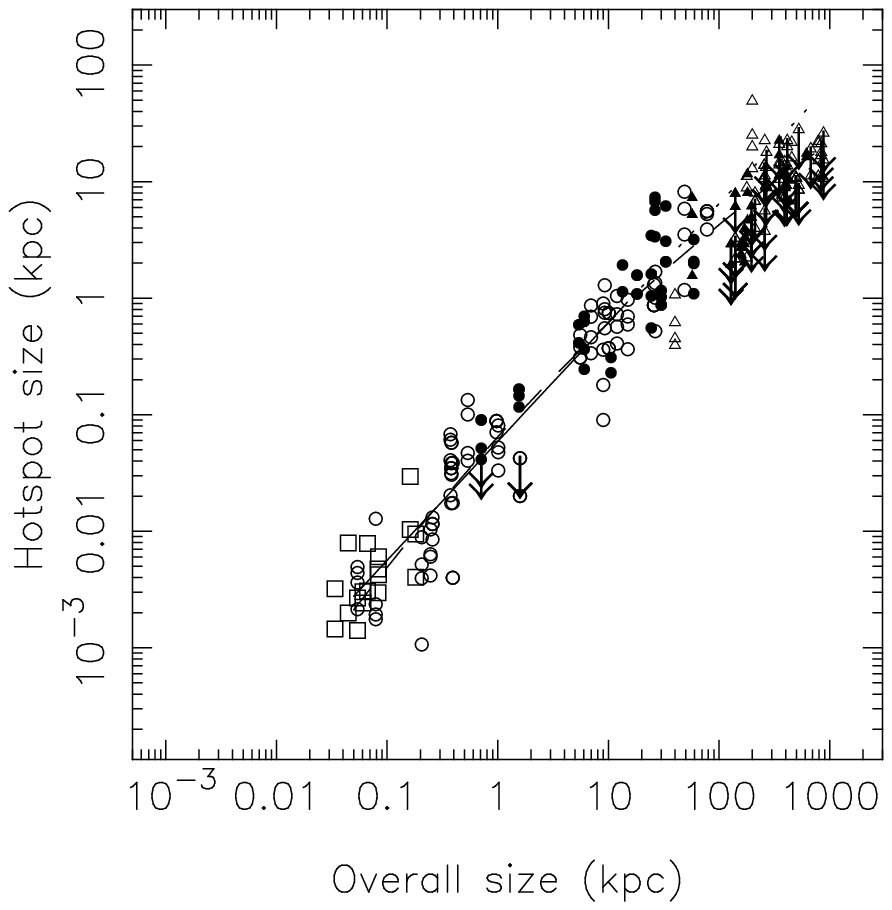,width=7.1in}
}
\vspace{-6.0in}
\hbox{
\hspace{-0.5in}
\psfig{figure=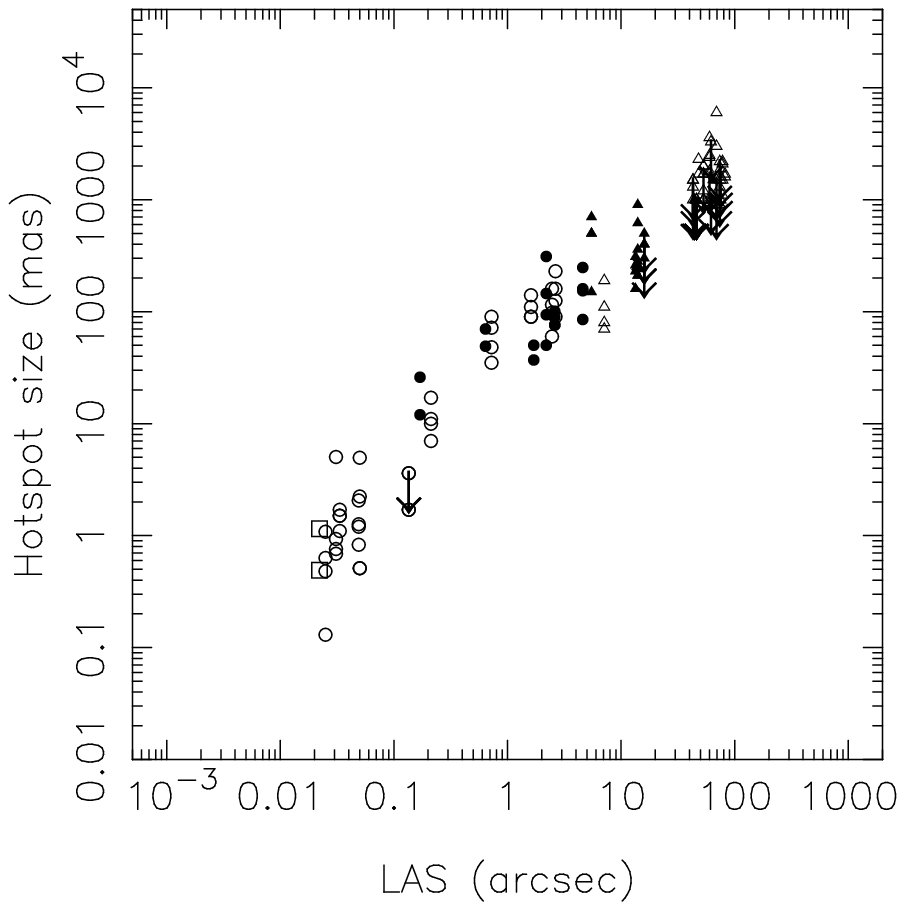,width=7.1in}
\hspace{-3.6in}
\psfig{figure=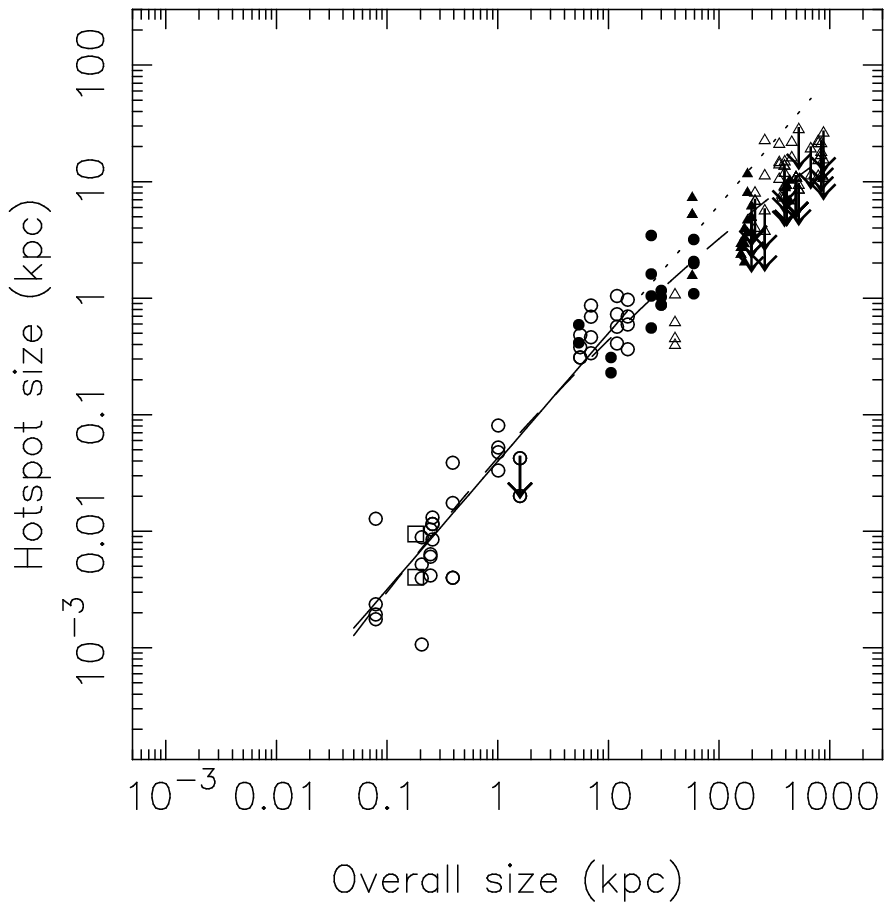,width=7.1in}
}
}
\vspace{-0.2in}
\caption{The size of each hotspot is plotted against the largest angular size of
each source defined to be the separation of the oppositely-directed hotspots.
Filled circles indicate quasars while open circles
denote radio galaxies from the sample of largely CSS and GPS objects (Table 1).
The knots in the jets of the compact symmetric objects are denoted by
open squares. The hotspot sizes of the 3CR sources described in the text (Table 2) are indicated by
open and filled triangles for the galaxies and quasars respectively.
The arrows indicate upper limits.
The upper panels indicate all sources observed with at least 10 resolution elements
along the main axes while the lower panels show those which have been observed with at least
20 resolution elements along the main axes.
The solid lines denote the linear least-squares fit to the average hotspot size
for CSS and GPS sources, defined to
be those below 20 kpc. These linear fits have been extended beyond 20 kpc by a dotted line
to highlight the flattening of the relationship for larger sources.
The dashed lines denote the parabolic fit to the average hotspot size of the
sources in both the samples. For sources
with upperlimits, the hotspot sizes have been estimated by assuming the true values to
be close to the limits.
}
\end{figure*}

\section{Collimation of radio jets}
In the upper-left panel of Figure 1, we plot the angular sizes of the major and minor
axes of each hotspot against the 
largest angular separation of each source in Tables 1 and 2, while 
on the upper-right panel we present the equivalent linear sizes for
the same sources. The lower panels show the same plots but only
for those sources which have been observed with at least about 20 resolution elements along
the long axis of the source.
Since we sometimes do not have information on the radio core, especially for the
CSS and GPS objects we have plotted the hotspot sizes against the overall
separation of the oppositely-directed hotspots. 
In this figure, we have also plotted the knots in the CSOs imaged by Taylor
et al. (1995, 1996). 
In this paper, we have made the least-squares fits to the hotspots 
by taking the average hotspot size for each source. This is defined to be the
geometric mean of each hotspot and, for those with hotspots on both sides, 
the average of the two oppositely-directed hotspots for each source. 
As mentioned earlier, a distance of 20 kpc was chosen as the canonical 
limit for the size of the CSS and GPS objects (cf. Fanti et al. 1990; O'Dea 1998).
A linear least-squares fit to the hotspots for the CSS and GPS objects in Figure 1 
observed with at least 10 resolution elements shows that if the jet width
is expressed as $d_{jet} = k l^{n}$, then  $n \sim 1.02\pm0.06$ and log$(k) \sim -1.23\pm0.05$.
For those observed with at least 20 resolution elements the value of 
$n$ is $1.10\pm0.07$ and log$(k) \sim -1.4\pm0.05$. The sizes of the knots in the CSOs
are consistent with this trend. 
There appears to be a well-defined relationship with 
the hotspot size increasing with the total size of the source upto a distance of about
20 kpc, the canonical limit for the sizes of CSSs. The sizes of the knots in the jets
are similar to those of the hotspots. This relationship is consistent with 
self-similar models for the evolution of radio sources during the CSS and GPS phase 
(cf. Kaiser \& Alexander 1997). 

\subsection{Comparison with larger 3CR sources observed with 
similar resolution elements}
To investigate this relationship for sources $>$ 20 kpc, 
we have compared the CSS and GPS sources (Table 1) 
with the  sample of 3CR sources (Table 2) which have been observed 
with similar resolution relative to the largest angular size as used for the 
CSS sources. The number of resolution elements, n$_b$, for the CSS 
sources vary from about 10 to 77 with a median value of about 19 while for 
the 3CR sources
n$_b$ ranges from about 10 to 39 with a median value of about 23. Only 5 of the 
CSSs have  n$_b$ $>$40. Considering the sources which have been observed with 
n$_b$ between about 20 and 40, the median values of n$_b$ for CSS and GPS, and 3CR
sources are 25 and 28 respectively, and a Kolmogorov-Smirnov test shows that 
the two distributions are the same with a probability of 66 per cent. 
The redshifts of the CSS and GPS sources range from about
0.1 to 2 with a median value of about 0.6 while the redshifts for the 3CR 
sample is somewhat larger ranging from about 0.16 to 2.0, with a median
value of about 0.8. 

The extension of the linear least-squares fit to the
CSS sources beyond 20 kpc shows that most of the larger sources 
observed with a similar value of n$_b$ lie below this line, 
suggesting a flattening of this relationship. This indicates recollimation
of the jets beyond $\sim$20 kpc.
Since most of the measured sizes are upper limits, it is difficult
to determine the precise nature of the relationship from the present
sample. Assuming that the sizes of the hotspots with upper limits are
close to these limits, we have attempted a linear least-squares as well
as a parabolic fit to the entire data. The parabolic fits are significantly
better, suggesting again a flattening of the relationship. To confirm the 
trend and determine the degree of flattening, we 
need to measure more precisely the sizes of the hotspots, rather than have limits
to their sizes. 

\subsection{3CR sources observed with larger values of n$_b$}
Over the last few years several authors have determined the sizes of the hotspots
of 3CR FRII sources from high-resolution VLA observations  
with n$_b$, the number of resolution elements along the axis of the source,
in the range of about 35 to 1040 (e.g. B94; F97; Black et al. 1992;
Leahy et al. 1997; Hardcastle et al. 1997). With higher values of n$_b$ one should
expect to see smaller-scale structures within the hotspots, and hence a decrease in
the size of the hotspots. A linear least-squares fit to
the hotspot size $-$ n$_b$ diagram for the B94, F97 and H98 samples 
exhibits a weak trend for the hotspot size to decrease with n$_b$ with a 
slope of $-0.15\pm0.10$. Because of different definitions of hotspots and also
difficulties in measuring their sizes, detailed comparisons are often difficult and
contentious. However, the sizes of the hotspots from the B94, F97 and H98 have
median values in the range of about 1.9 to 4.5 kpc and are consistent with a flattening of
the hotspot size $-$ linear size relationship beyond about 20 kpc. The knots in the
jets (Table 9 of B94) are also consistent with this flattening.

\begin{figure}
\vspace{-5.8in}
\hbox{
\hspace{-0.5in}
\psfig{figure=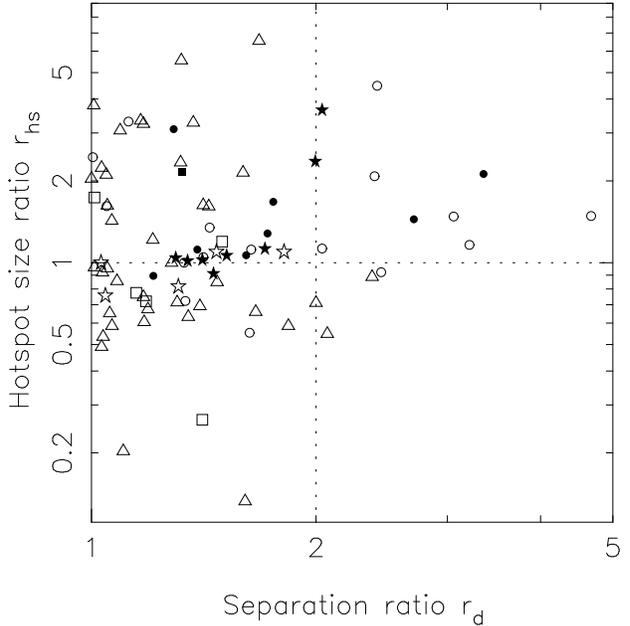,width=7.1in}
}

\vspace{-0.2in}
\caption{The ratio, r$_{hs}$, of the size of the hotspot farther from the radio core to
that of the oppositely-directed one closer to the core, is plotted against the 
separation ratio, r$_d$, defined to be the ratio of the separations of the corresponding
hotspots from the core. The open and filled circles denote galaxies and quasars from
the sample of largely CSS and GPS objects (Table 1), the open and filled stars denote
galaxies and quasars from B94 and F97, while the triangles represent radio galaxies from
H98. Filled and open squares indicate quasars and radio galaxies from Table 2 which
have not been already plotted.
}
\end{figure}

\begin{figure}
\vspace{-5.5in}
\hbox{
\hspace{-0.5in}
\psfig{figure=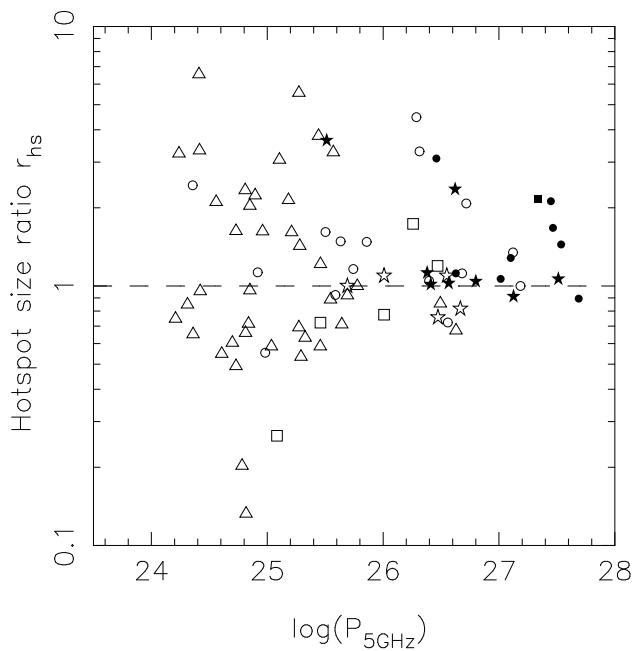,width=7.1in}
}
\vspace{-0.2in}
\caption{The ratio, r$_{hs}$, of the size of the hotspot farther from the radio core to
that of the oppositely-directed one closer to the core, is plotted against the 
total luminosity at 5 GHz. The symbols are as defined in Figure 2. 
}
\end{figure}

\subsection{Opening angle of the jets}
Using the geometric mean of the major and minor axes for each hotspot and the average 
of the hotspots for each source, we estimate the mean opening angle, ($\phi_{obs}$),
for sources below 20 kpc to be about 7$^\circ$. For sources observed with
atleast 20 beam widths along the source axes the mean opening angle is about 5$^\circ$.
The estimates of the mean opening angle is roughly consistent with estimates for 
individual jets in different angular and linear scales. It is relevant to note
that the opening angle 
estimated from the hotspots could, in principle, be larger than those estimated 
from knots in the jets because of the dentist drill effect (Scheuer 1982; Norman 1996). In
the radio galaxy Cygnus A, the opening angle of the jet determined from the knots
as well as the low surface brightness inter-knot regions is $\approx$1.6$^\circ$
(Carilli et al. 1996), although the morphology of the jet is consistent with
regions of larger opening angle of $\approx$5$^\circ$ (Perley, Dreher \& Cowan 1984).
The jet in M87 is well-collimated on all scales out to about 2 kpc. The 
innermost VLBI structure indicates an opening angle of 18$^\circ$, while VLA
observations indicate that between the jet and knot A, the jet appears as
a uniformly expanding cone with an opening angle of 7$^\circ$ (Biretta 1996;
Biretta \& Junor 1995). Among the core-dominated radio sources, one of the
best studied ones is the quasar 3C345 which has an opening angle of 26$^\circ$
(Zensus, Cohen \& Unwin 1995). Other examples are NRAO 140 with an opening 
angle of 12$^\circ$ (Marscher 1988; Unwin \& Wehrle 1992) and 0836+710 
which has a mean opening angle of 4.4$^\circ$ (Hummel et al. 1992).

\subsection{Asymmetry in collimation}
We have also examined the dependence of the ratio of the sizes of the two 
oppositely-directed hotspots, r$_{hs}$, on their relative separations from the nucleus. 
In Figure 2 we plot the ratio, r$_{hs}$,
defined to be the ratio of the size of the farther hotspot to that of the nearer one, against
the separation ratio, r$_d$, for all the sources with 
a detected radio core and hotspots on both sides. The separation ratio, r$_d$,  is 
defined to be $\geq$ 1, while 
r$_{hs}$ is the ratio of the geometric mean of the corresponding hotspots. We have
used only those hotspots without an upperlimit to any of their axes and with a detected
radio core from the sample of largely CSS and GPS objects (Table 1), the 3CR objects
listed in Table 2 and the observations of B94, F97 and H98. 

For the 33 high-luminosity sources with a radio luminosity at 5 GHz
$\geq 10^{26}$ W Hz$^{-1}$ sr$^{-1}$, 24 have r$_{hs}$ $>$ 1, implying that
the farther hotspot is larger. This trend is 
also seen in the most asymmetric objects, defined to be those with r$_d >$ 2.
Of the 12 sources with r$_d >$ 2, 9 have r$_{hs}$ $>$ 1, and almost all these objects are CSSs.
Either light-travel time effects or an intrinsic or environmental asymmetry can 
produce the effect in the observed sense. 
For a source inclined at about 50$^\circ$ to the
line-of-sight and a hotspot speed of about 0.5c, r$_d$ is close to about 2.
There have been several estimates of hotspot advance speeds in VLBI-scale double sources. 
O'Dea (1998) lists 7 objects which have upper limits to component proper motion
which are subluminal, ranging from 0.05-0.5c. Since O'Dea's review the upper limit
for 1934$-$638 has been revised to $\sim$0.03$\pm$0.2c (Tzioumis et al. 1998), and
an upper limit for 1607+26 (CTD93) has been reported to be less than 0.35c (Shaffer \&
Kellermann 1998). 
Current estimates of hotspot speeds suggest that the effects of an asymmetry dominate.
Almost all these objects with r$_d >$2 are compact steep spectrum radio sources, and the
deficit of objects with r$_d >$2 and r$_{hs} <1$ is possibly a reflection of their 
evolution in an asymmetric environment.

The tendency for r$_{hs}$ $>$ 1 is not seen in objects of lower luminosity (Figure 3). For example,
in the 50 sources with radio luminosity at 5 GHz less than  10$^{26}$ W Hz$^{-1}$ sr$^{-1}$,
only 24 have r$_{hs}$ $>$ 1. A number of these objects have  r$_{hs}$ significantly smaller
than 1, which could be due to an intrinsic asymmetry in the collimation of jets on 
opposite sides of the nucleus.

\section{Discussion and concluding remarks}
We summarise the principal trends reported in this paper.

\begin{enumerate}

\item The relationship between the hotspot sizes and the overall size of
the CSS and GPS sources and studies of individual knots in the jets in 
these sources, suggest that they evolve in an approximately self-similar way. 
The hotspot size increases with distance from the core as $d_{jet} 
\propto l^{n}$ where 
$n \sim$ 1.0 for sources smaller than about 20 kpc. This is similar for sources
which have been observed with at least either 10 or 20 resolution elements, n$_b$, along 
the main axes so that the hotspots could be identified reliably. 

\item For larger sources observed with at least either 10 or 20 resolution elements along the
main axes and with a similar distribution of n$_b$ to the sample of largely
CSS and GPS sources, there appears to be a
flattening of the relationship. However, since most of the sources have upper limits
to their hotspot sizes the precise slope could not be determined reliably. 

\item For samples of sources observed by Bridle et al. (1994) and Fernini et al. (1993, 1997)
with n$_b$ in the range of about 40 to 200, and a sample of 3CR sources with z$<$0.3 summarized by 
Hardcastle et al. (1998) with n$_b$ in the range of about 35 to 1040, the dependence of 
hotspot size on n$_b$ is weak. The hotspot size varies by a factor of about 2 for an
increase in n$_b$ by a factor of about 300. The hotspot sizes of these large sources
are consistent with a flattening of the hotspot size-linear size relationship 
seen for CSS and GPS objects. 

\item There is a tendency for the farther hotspot to be larger in sources with a luminosity at
5 GHz $>$ 10$^{26}$ W Hz$^{-1}$ sr$^{-1}$. 
This could be caused by both light travel time effects as well as an asymmetric environment.
No such trend is seen in objects of lower luminosity where the nearer hotspot is significantly larger in 
a number of sources. This could be due to an asymmetry in collimation on opposite sides of the nucleus.

\item There is a trend for the farther hotspot to be larger
in the most asymmetric objects, defined to be those with r$_d >$ 2.
Almost all the objects with r$_d >$ 2 are CSS objects. Current estimates of hotspot advance
speeds suggest that these trends are due to an intrinsic asymmetry in the environment 
rather than light travel time effects. 

\end{enumerate}

In the CSS  phase, the variation of the sizes of the hotspots and
the widths of the knots in the jets with linear size 
could be due to the ambient pressure falling with distance from the nucleus,
if the jet is pressure confined.
The confinement of the jet by the ambient medium depends on
whether the jet pressure is comparable to the external pressure.
The equipartition pressure in the hotspots or knots  varies from about 10$^{-6}$ N m$^{-2}$
on scales of a few tens of parsec to about
$10^{-9}$ N m$^{-2}$ at about 10 kpc from the nucleus (Fanti et al. 1995; Readhead 1995;
Readhead et al. 1996a,b). For large sources, the equipartition pressure in the knots of the jets
is in the range of about $10^{-10}$ to  $10^{-12}$ N m$^{-2}$ (Potash and Wardle 1980;  Bridle et al. 1994).
Although the nature of the confining medium within about 20 kpc, i.e. the CSS phase,
needs to be better understood (cf. Fanti et al. 1995), the gaseous components seen in
absorption at X-ray, UV and optical wavelengths (Mathur et al. 1994; Elvis et al. 1996;
Netzer 1996) could play an important role in addition to the broad-line and narrow-line gas.

For the larger FRII sources the flatter relationship 
seen in the present data suggests recollimation of the jets beyond the CSS phase. 
It is important to understand the physical process responsible for the
recollimation of jets. Two interesting scenarios
which have suggested for recollimation are 
hydrodynamic collimation involving shocks (Sanders 1983; Falle \& Wilson 1985;
Komissarov \& Falle 1996)  and magnetic collimation (Begelman, Blandford \& Rees 1984; 
Appl \& Camenzind 1992, 1993a,b; Appl 1996). The latter could provide a natural explanation
of a constant width in large sources, which appears to be suggested by some of the 
observations such as those of B94 and F97, and in 3C353 (cf. Swain et al. 1996).

\section*{Acknowledgments}
We thank Robert Laing, Peter Scheuer and an anonymous referee for their critical comments and
suggestions, and our colleagues at NCRA 
especially Gopal-Krishna and late Vijay Kapahi for 
their comments on this piece of work, and Judith Irwin, Ishwara-Chandra and  
Vasant Kulkarni for their detailed comments on the manuscript. This research has made use of the 
NASA/IPAC extragalactic database (NED)
which is operated by the Jet Propulsion Laboratory, Caltech, under contract
with the National Aeronautics and Space Administration. 

{}

\end{document}